\documentclass[twocolumn,preprintnumbers,nobibnotes,floatfix,superscriptaddress]{revtex4}
\usepackage{graphicx,color}
\usepackage{amssymb}
\usepackage{amsmath}
\usepackage{subfigure}
\usepackage{float}
\usepackage{bm}
\usepackage[T1]{fontenc} 
\usepackage{array}
\usepackage{hyperref}

\def\simge{\mathrel{%
       \rlap{\raise 0.511ex \hbox{$>$}}{\lower 0.511ex \hbox{$\sim$}}}}
\def\simle{\mathrel{
       \rlap{\raise 0.511ex \hbox{$<$}}{\lower 0.511ex \hbox{$\sim$}}}}

\makeatletter
\newcommand{\cpn}{CP$^{N\!-\!1}$\ }
\newcommand*{\doi}[1]{}
\newcommand{\figcaption}[1]{\def\@captype{figure}\caption{#1}}
\newcommand{\tblcaption}[1]{\def\@captype{table}\caption{#1}}

\newcommand{\la}{\langle\,}
\newcommand{\ra}{\,\rangle}

\newcommand{\vsub}{V_{\rm sub}}
\newcommand{\avsub}{V_{\rm sub}/a^4}
\newcommand{\qsub}{Q_{\rm sub}}
\newcommand{\fsub}{f_{\rm sub}}
\newcommand{\nape}{n_{\rm APE}}

\newcommand{\SU}{$\mathrm{SU}$}

\renewcommand{\th}{\theta}

\makeatother

\begin{document}
\title{Peeking into the $\th$ vacuum}

\def\KEK{High Energy Accelerator Research Organization (KEK), Tsukuba 305-0801, Japan}
\def\SOKENDAI{Graduate University for Advanced Studies (SOKENDAI), %
Tsukuba 305-0801, Japan}

\author{Ryuichiro Kitano}
\affiliation{\KEK}
\affiliation{\SOKENDAI}

\author{Ryutaro Matsudo}
\affiliation{\KEK}

\author{Norikazu Yamada}
\affiliation{\KEK}
\affiliation{\SOKENDAI}

\author{Masahito Yamazaki}
\affiliation{
Kavli Institute for the Physics and Mathematics of the Universe
(WPI), University of Tokyo, Kashiwa, Chiba 277-8583, Japan}

\date{\today}

\begin{abstract}
 We propose a subvolume method to study the $\th$ dependence of the free
 energy density of the four-dimensional \SU($N$) Yang-Mills theory on
 the lattice.
 As an attempt, the method is first applied to \SU(2) Yang-Mills theory at
 $T=1.2\,T_c$ to understand the systematics of the method.
 We then proceed to the calculation of the vacuum energy density and
 obtain the $\th$ dependence qualitatively different from the high
 temperature case.
 The numerical results combined with the theoretical requirements
 provide the evidence for the spontaneous CP violation at
 $\theta = \pi$, which is in accordance with the large $N$ prediction
 and indicates that the similarity between 4d SU($N$) and 2d \cpn
 theories does not hold for $N$=2.
\end{abstract}

\maketitle

\section{Introduction}\label{sec:intrroduction}

The $\theta$ parameter of the 4d Yang-Mills theory controls relative
weights of different topological sectors in the path integral.
Despite long history, it still remains as a challenging problem to
identify the effect of the $\theta$ parameter on the non-perturbative
dynamics of the theory.

For the special value $\theta=\pi$ the Lagrangian has CP symmetry, and
we can ask whether or not the CP symmetry is spontaneously broken.
In the large $N$ limit~\cite{tHooft:1973alw} spontaneous CP violation at
$\theta=\pi$ was demonstrated in
Refs.~\cite{Witten:1980sp,tHooft:1981bkw,Witten:1998uka}.
For finite $N$ a mixed anomaly between the CP symmetry and the
$\mathbb{Z}_N$ center symmetry shows that the CP symmetry in the
confining phase has to be broken~\cite{Gaiotto:2017yup}.
A similar conclusion was derived by studying restoration of the
equivalence of local observables in \SU($N$) and \SU($N$)/$\mathbb{Z}_N$
gauge theories in the infinite volume limit~\cite{Kitano:2017jng}.
(See also Refs.~\cite{Azcoiti:2003ai,Yamazaki:2017dra,Wan:2018zql}.)
While these theoretical developments have narrowed down possible
scenarios, an explicit nonperturbative calculation is necessary to
unambiguously settle the fate of the CP symmetry at $\theta=\pi$.
Any direct numerical simulation at $\theta=\pi$, however, has been
difficult due to the notorious sign problem~\footnote{For recent related
efforts towards direct simulations, see, for example,
Refs.~\cite{Hirasawa:2020bnl,Gattringer:2020mbf}.}.
 
In Ref.~\cite{Kitano:2020mfk} three of the authors of the present paper
studied the vacuum energy density of the 4d \SU(2) Yang-Mills theory
around $\theta=0$ by lattice numerical simulations. The case of \SU(2)
gauge group is of particular interest since $N=2$ is farthest away from
the large $N$ limit: there is a well-known parallel between 2d CP$^1$
and 4d \SU(2) model~\footnote{See Refs.~\cite{Yamazaki:2017ulc,
Yamazaki:2017dra} for more precise connections between the two.}, and
the known vacuum at $\th=\pi$ in the former \cite{CP1} alludes to the
appearance of gapless theory in the latter.
By observing that the first two numerical coefficients in the $\theta$
expansion obey the large $N$ scaling~\footnote{See
Refs.~\cite{Lucini:2001ej,DelDebbio:2002xa,Bonati:2016tvi} for large $N$
scaling of the first two coefficients in the $\theta$ expansion in the
SU($N$) gauge theory.
For SU(2) theory the first coefficient in the $\theta$ expansion, the
topological susceptibility $\chi$, was estimated in
Refs.~\cite{deForcrand:1997esx,Alles:1997qe,DeGrand:1997gu,Lucini:2001ej,Berg:2017tqu}.},
it was inferred in  Ref.~\cite{Kitano:2020mfk} that the 4d \SU(2)
Yang-Mills theory at $\theta=\pi$ has spontaneous CP breaking, contrary
to the naive expectation from the 2d CP$^1$ model.

In this work, we develop a subvolume method to explore the $\theta$
dependence of the free energy \emph{without any series expansion in
$\theta$} and apply to 4d \SU(2) theory.
We find that it indeed works and show an evidence of spontaneous CP
violation at $\theta=\pi$ at zero temperature, in consistency with the
results of Ref.~\cite{Kitano:2020mfk}.
The subvolume method here is inspired by Ref.~\cite{Luscher:1978rn} and
similar to that introduced in Ref.~\cite{KeithHynes:2008rw} to study 2d
\cpn model.

\section{Subvolume Method and Lattice Set Up}
\label{sec:method}

In what follows, the subvolume method is described.
After generating a number of gauge configurations at $\th=0$ and
implementing the $n_{\rm APE}$ steps of APE
smearing~\cite{Albanese:1987ds}, the topological charge density $q(x)$
is calculated with the five-loop improved topological charge
operator~\cite{deForcrand:1997esx} on the lattice.
Defining the subvolume topological charge $\qsub$,
\begin{align}
& \qsub = \displaystyle\sum_{x\in \vsub} q(x)\ ,
\end{align}
we calculate the free energy density for the subvolume
as~\cite{Luscher:1978rn,KeithHynes:2008rw},
\begin{align}
& f_{\rm sub}(\th)
= \frac{-1}{\vsub}\ln \frac{Z(\th)}{Z(0)}
= \frac{-1}{\vsub}\ln \,\la \cos\left(\theta\,Q_{\rm sub}\right)\ra
\ ,
\label{eq:fL}\\
& Z(\th)=\int\!\!{\cal D}U\,e^{-S_g+i\th\qsub}
\ ,
\end{align}
where $U$ denotes the link variable and $S_g$ the gauge action.
Note that the expectation value $\langle\cdots\rangle$ is estimated on
the $\th=0$ configurations.
The free energy density is then obtained as the infinite volume
limit of $f_{\rm sub}(\th)$:
\begin{align}
& f(\th)
= \lim_{\vsub\to\infty} \fsub(\th)\ .
\label{eq:fsub}
\end{align}
The goal of this study is to answer whether the spontaneous CP
violation does occur at $\theta=\pi$.
Thus, the order parameter should be useful and is calculated through
\begin{align}
& \frac{d\,f(\th)}{d\,\th}
= \lim_{\vsub\rightarrow\infty} 
  \frac{d\,f_{\rm sub}(\th)}{d\,\th}\ ,
\label{eq:dfdth}\\
& \frac{d\,f_{\rm sub}(\th)}{d\,\th}
= \frac{1}{\vsub}
  \frac{\left\langle \qsub\, \sin(\th\qsub) \right\rangle}
         {\la\cos(\th \qsub)\ra}\, .
\label{eq:dfdthL}
\end{align}
$f$ and $df/d\th$ are evaluated separately and later used to make a
consistency check.

Some cautions are in order.
First, the size of the subvolume $\vsub$ cannot be arbitrary.
Obviously it has to be large enough to cover the typical correlation
length of the system, which is considered to be around
$1/(aT_c)= 9.50$~\cite{Giudice:2017dor} in the lattice unit.
We thus restrict the size of the subvolume to $\vsub \ge 10^4$.
Suppose that $\vsub$ is large enough but $\vsub\ll V_{\rm full}$, where
$V_{\rm full}$ denotes the full volume.
Then, the resulting $f(\theta)$ is expected to be independent of
$V_{\rm full}$, and $\fsub$ would shows a scaling behavior as a function
of $\vsub$.
As $\vsub$ grows and approaches $V_{\rm full}$, $\qsub$ becomes close to
an integer global $Q$, and $\fsub$ starts to converge a full volume
result, in which we are not interested as it becomes clear later.
The critical goal in this method is to identify the scaling behavior
from which the infinite volume limit is extracted.

Secondly, the observables \eqref{eq:fL} and \eqref{eq:dfdthL} contain
the expectation value $\langle\cos(\th \qsub)\rangle$.
Either observable becomes incalculable when
$\langle\cos(\th \qsub)\rangle$ fluctuates across zero.
This is the sign problem in this method and sets the $\th$ dependent
upper limit on the size of $\vsub$.
Thus, another crucial point is whether the scaling behavior is
realized before $\vsub$ reaches the upper limit.
We will return to this point in the discussion section.

The free energy density of 4d \SU(2) Yang-Mills theory is calculated at
high and zero temperatures below.
We employ the configurations generated in our previous work at
$T=0$~\cite{Kitano:2020mfk}, while the ensemble corresponding to
$T=1.2\, T_c$ is newly generated at the lattice coupling $\beta=1.975$,
the same as the $T=0$ case.
The simulation parameters are summarized in Tab.~\ref{tab:lat-sim}.
\begin{table}[t]
 \begin{center}
 \begin{tabular}{|c|c|c|c|c|}
 \hline
  $N_S^3 \times N_T$ & $T/T_c$ &
  [${\nape}_{\rm min},\ {\nape}_{\rm max}$] 
& statistics & $a^4\chi \times 10^5$\\
 \hline
  $24^3\times  8$ & 1.21 & [25,\ 45] & 10,000 & 1.35(5) \\
  $24^3\times 48$ & 0    & [20,\ 40] & 68,000 & 2.54(3) \\
 \hline
 \end{tabular}
 \caption{
  The lattice parameters of the ensembles.
  The tree-level Symanzik improved action~\cite{Weisz:1982zw} is adopted
  with the periodic boundary conditions in all four directions.
  The value $1/(aT_c)=9.50$~\cite{Giudice:2017dor} gives the value of
  $T/T_c$ in the table, where $T_c$ is the critical temperature at
  $\th=0$.
  }
 \label{tab:lat-sim}
 \end{center}
\end{table}

The topological charge density measured at each smearing step is
uniformly shifted as $q(x)\to q(x)+\epsilon$ at every configurations so
that the global topological charge $Q=\sum_{x\in V_{\rm full}}q(x)$
takes the integer closest to the original value.
The calculation of $\qsub$ is carried out every five smearing steps.
For the APE smearing, we take $\alpha=0.6$ in the notation of
Ref.~\cite{Alexandrou:2017hqw}.

Topological observables on the lattice can be distorted by topological
lumps originating from lattice artifacts.
One can take away those by the smearing procedure, but at the same time
the smearing may deform physical topological excitations, too.
We studied this point in detail in Ref.~\cite{Kitano:2020mfk}, and
developed the procedure to restore the physical information.
The procedure consists of the extrapolation of the observables to the
zero smearing limit by fitting over a suitable interval of the smearing
steps.
The fit range is fixed in advance by examining the response of the
global topological charge to the smearing as done in
Ref.~\cite{Kitano:2020mfk}.
The resulting fit ranges and the topological susceptibilities
$\chi=\la Q^2 \ra/V_{\rm full}$ thus determined using the global
topological charge $Q$ are shown in Tab.~\ref{tab:lat-sim} for later
use.

The $\theta$-dependence is explored in the range of $\th=k\,\pi/10$ with
$k\in [1,\,20]$.
Each configuration is separated by ten Hybrid Monte Carlo (HMC) trajectories.
In the following analysis, statistical errors are estimated by the
single-elimination jack-knife method with the bin size of 500 and 100
configurations for zero and high temperatures, respectively.
We mainly show the analysis of $f(\theta)$, but $df(\theta)/d\theta$ is
analyzed in parallel with similar quality.

\section{Testing the method at High Temperature}
\label{sec:finiteT}

We first apply the subvolume method to the calculation of the free
energy density above $T_c$, where the instanton prediction,
$f(\th)\sim\chi(1-\cos\th)$~\cite{'tHooft:1976fv,Callan:1977gz,Gross:1980br},
is believed to be valid and numerically supported for \SU($N$) with
$N\ge 3$~\cite{Bonati:2013tt,Frison:2016vuc} as well.
Using the translational invariance, a single-sized subvolume
$\avsub=l^3\times N_T$ with $l=10,\ 12,\ \cdots,\ 24$ is taken from 64
places per a configuration, and the results are averaged.

\begin{figure}[tb]
  \begin{center}
  \includegraphics[width=0.5 \textwidth]
  {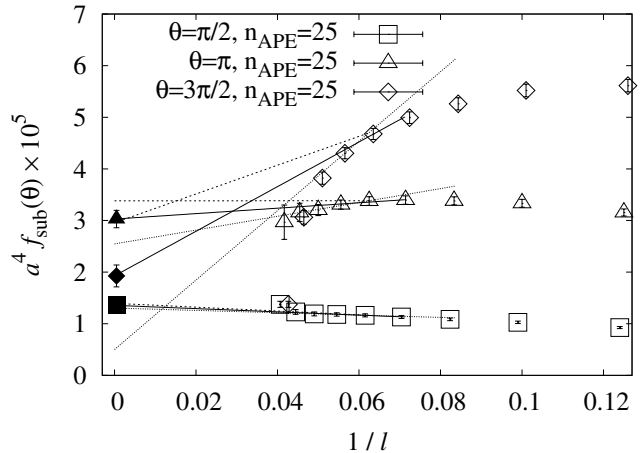}
  \end{center}
 \caption{
 The subvolume dependence of $f_{\rm sub}(\th)$ and the extrapolation to
 the infinite volume limit.
 }
 \label{fig:v-vs-f-24x8}
\end{figure}

The $l$ dependence of $\fsub$ is shown in Fig.~\ref{fig:v-vs-f-24x8}.
It is found that in general the measured $l$ dependence is not constant
and the leading correction linear in $1/l$ is inevitable.
The linear dependence is seen for $\th=\pi/2$ for $l\le 22$, whereas it
ends around $l=20$ for $\th=\pi$ and it is hard to determine the linear
region for $\th=3\pi/2$.
Thus, it turns out to be difficult to identify the scaling region
unambiguously especially when $\th$ is large, and hence we give up the
precise determination.
Instead, we choose three fit ranges in each extrapolation and try to
estimate the potential size of the systematic uncertainty due to the
ambiguity of the scaling region.
Three fit ranges, $l\in [12,16]$, $[14,18]$ and $[16,20]$, are examined
when fitting to the expected scaling behavior
\begin{align}
 f_{\rm sub}(\th) = f(\th) + \frac{a^{-1}\,s(\th)}{l}\, ,
\label{eq:v-fit-form}
\end{align}
where $s(\th)$ denotes the surface tension of the nonzero $\th$ domain
and $a$ the lattice spacing.
All fits performed in this analysis yield $\chi^2/\mbox{dof} < 3$.
It is interesting to see that the relative relation
$\fsub(3\pi/2)>\fsub(\pi)$ at small $l$ flips toward the large
$l$ limit and $f(\th)$ ends up with non-monotonic function.
Since the data of $\fsub$ smoothly (but sometimes rapidly) connect the
full volume ($l^3\times 6=24^3\times 6$) results from the above, the
extrapolation fitting the data near the full volume tends to be smaller
than that fitting the data far from the full volume.
As a result, the discrepancy, {\it i.e.} the potential size of the
systematic uncertainty, turns out to be larger at larger $\th$.

\begin{figure}[tb]
 \centering
  \includegraphics[width=0.5 \textwidth]
  {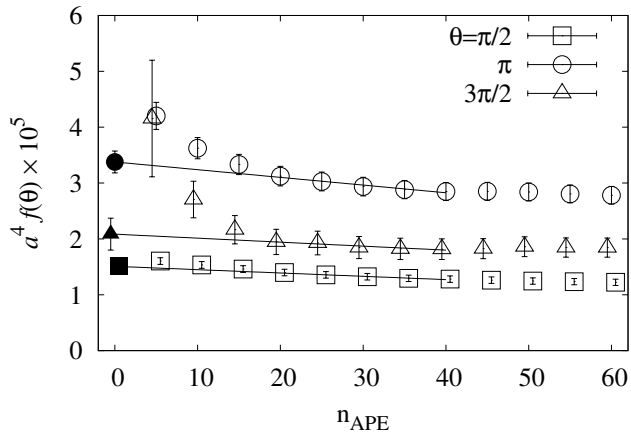}
 \caption{
 The linear extrapolation of $f(\th)$ to $\nape=0$, where $f(\th)$ is
 obtained by the fit with $l\in[14,18]$.
 }
 \label{fig:nape-vs-f-24x8}
\end{figure}

The results thus obtained are then extrapolated to $\nape=0$ at each
value of $\th$ with the fit range shown in Tab.~\ref{tab:lat-sim}.
In the extrapolation, the linear fit goes well with
$\chi^2/\mbox{dof}<3$.
The stability against small shifts of the fit range is seen in
Fig.~\ref{fig:nape-vs-f-24x8}.

\begin{figure}[tb]
 \centering
  \includegraphics[width=0.5 \textwidth]
  {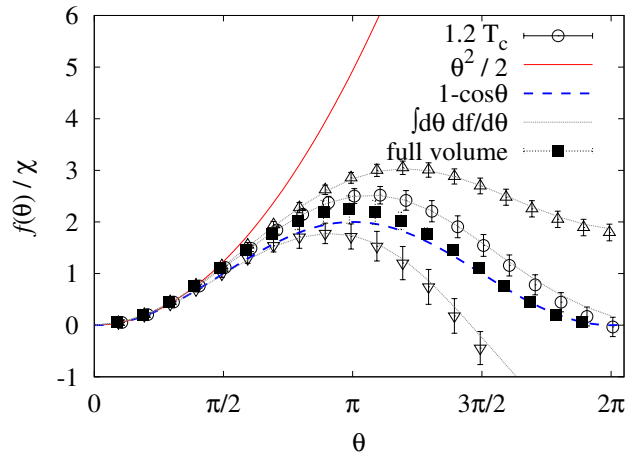}
 \caption{
 The $\theta$ dependence of $f(\th)$ at $T=1.2\,T_c$.
 The results obtained with different fit ranges in
 Fig.~\ref{fig:v-vs-f-24x8}, $l\in [12,16]$, $[14,18]$ and $[16,20]$,
 are shown in triangle-up, circle and triangle-down, respectively.
 }
 \label{fig:th-vs-f-24x8}
\end{figure}

Finally, the free energy density obtained with three fit ranges are
shown in Fig.~\ref{fig:th-vs-f-24x8} together with the full volume
result (filled squares), where $f(\th)$ is normalized by the topological
susceptibility in Tab.~\ref{tab:lat-sim}.
The prediction from the dilute instanton gas approximation,
$1-\cos(\th)$, is shown by the dashed curve.
The function, $\th^2/2$, is also shown as the solid curve for
comparison.
Taking into account the uncertainty arising from the ambiguity of the
scaling region, the numerical results are consistent with the
instanton prediction.
Note that non-monotonic behavior of $f(\th)$ seems robust at high
temperature but is far from obvious before the extrapolations, as the
surface tension term in Eq.~\eqref{eq:v-fit-form} is monotonic.

$f(\th)$ can also be obtained from the numerical integration of
$df(\th)/d\th$ as shown by the dotted curves.
The agreement with those curves supports that the two nontrivial
extrapolations included in the whole analysis do not pick up accidental
fluctuations and are stable.

The result with full volume is found to well agree with the instanton
prediction.
One may think that this is the simplest way to obtain $f(\th)$.
However, we will see that it does not work at $T=0$.
From the test, assuming that the instanton prediction is valid at high
temperature, we learn that the scaling behavior of $\fsub$ would be
linear and the region showing such a behavior starts around the
dynamical length scale ($\sim 1/(aT_c)$).

\section{Applying to Zero Temperature}
\label{sec:zeroT}

\begin{figure}[tb]
 \centering
  \includegraphics[width=0.5 \textwidth]
  {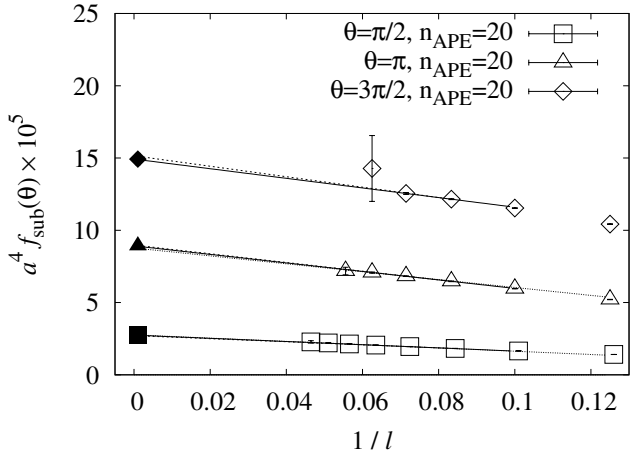}
 \caption{
 The linear extrapolation of $f_{\rm sub}(\th)$
 to the infinite volume limit.
 }
 \label{fig:v-vs-f-24x48}
\end{figure}

Next we apply the subvolume method to calculate the vacuum energy
density.
This time the subvolume is defined by $\avsub=l^4$ with 
$l=10, 12,\ \cdots,\ 24$ and taken from 512 places per configuration.
The $l$ dependence of $\fsub(\th)$ is shown in
Fig.~\ref{fig:v-vs-f-24x48} as before.
Due to the sign problem in this method, some results at large $\th$ and
large $l$ could not be calculated.
But the available data show linear behavior.
Following the previous analysis, three fit ranges of $l\in [10,14]$,
$[12,16]$ and $[14,18]$ are taken in the fit to \eqref{eq:v-fit-form} to
estimate the systematic uncertainty.
Contrary to the high temperature case, $f(\th)$ turns out to be stable
against the variation of the fit range, and does not show any sign of
the flip, indicating monotonic behaviors of $f(\th)$ as a function of
$\th$.

\begin{figure}[bt]
 \centering
  \includegraphics[width=0.5 \textwidth]
  {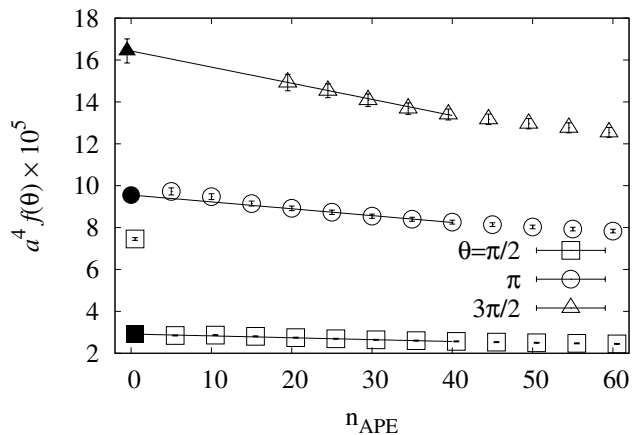}
 \caption{
 The linear extrapolation of $a^4f(\th)$
 to $\nape=0$ for the data obtained with $l\in[12,16]$.
 }
 \label{fig:nape-vs-f-24x48}
\end{figure}

The linear extrapolation to $\nape=0$ is carried out with the fit range
shown in Tab.~\ref{tab:lat-sim}, and the fit is found to work well
with $\chi^2/$dof $<3$ as shown in Fig.~\ref{fig:nape-vs-f-24x48}.
The stability against shift of the fit range is also confirmed.

\begin{figure}[tb]
 \centering
  \includegraphics[width=0.5 \textwidth]
  {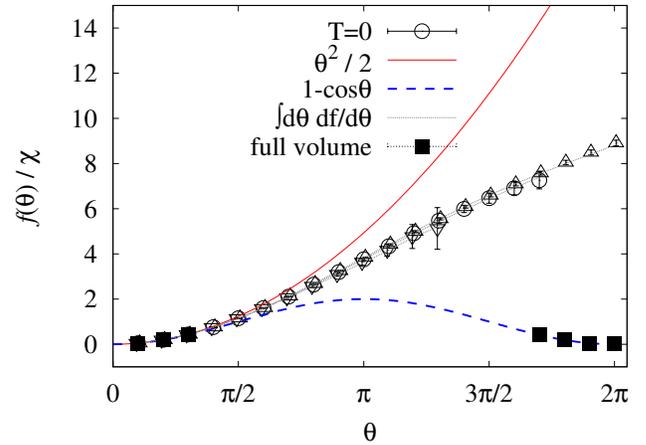} \\
  \includegraphics[width=0.5 \textwidth]
  {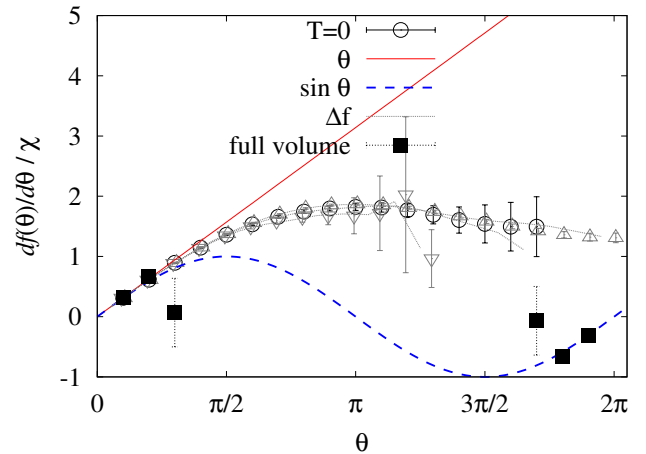}
 \caption{
 $f(\th)$ (top) and $df(\th)/d\th$ (bottom).
 The results obtained with different fit ranges in
 Fig.~\ref{fig:v-vs-f-24x48}, $l\in [10,14]$, $[12,16]$ and $[14,18]$,
 are shown in triangle-up, circle and triangle-down, respectively.
 }
 \label{fig:th-vs-f-24x48-f-df}
\end{figure}

Finally, the resulting $f(\th)$ and $df(\th)/d\th$ are shown in
Fig.~\ref{fig:th-vs-f-24x48-f-df} together with the predictions from the
large $N$ ($\th^2/2$) and the instanton calculus ($1-\cos\th$).
The stability of the two extrapolations during the analysis is confirmed
as $f(\th)$ and $df(\th)/d\th$ well agree with the dotted curves.
While the full volume calculation works only in the vicinity of $\th=0$,
the subvolume method succeeds to calculate, at least, to $\th\sim \pi$.
There are crucial differences from the high temperature case.
First, the different choices of the fit range in $l$ yield consistent
results, and hence the potential systematic error from the ambiguity of
the scaling region seems to be under control.
Second, $f(\th)$ is a monotonically increasing function, at least, to
$\th\sim\pi$, and the direct calculation of $df(\th)/d\th$ clearly shows
$d\,f(\th)/d\th|_{\th=\pi} \ne 0$.
Since $d\,f(\th)/d\th = -i \langle q(x) \rangle$ is CP odd, we conclude
that CP is spontaneously broken at $\theta = \pi$ in the vacuum of the
4d \SU(2) Yang-Mills theory~\footnote{See also Refs.~\cite{Unsal:2012zj}
and \cite{Unsal:2020yeh} for analytic discussions.} and that there is a
phase transition to recover the CP symmetry at some finite temperature.
In other words, it is found that the 4d SU(2) Yang-Mills theory is in the
large-$N$ class unlike the 2d CP$^1$ model~\footnote{There is a logical
possibility that there is a phase transition at some $\theta$ below
$\pi$, which the subvolume method could not detect.
In that case, the CP symmetry may be left unbroken at $\theta = \pi$.
In any case, the fact that the free energy does not show the $2\pi$
periodicity indicates that there are multiple branches in the vacuum
structure as in the case of the large $N$ limit.
We thank Yuya Tanizaki for discussion on this point.}.

\section{Discussion}
\label{sec:discussion}

The symmetry of \SU($N$) gauge theories indicates $f(\th)=f(-\th)$ and
$f(\th)=f(\th+2\pi)$.
In the subvolume method, $f(\th)=f(-\th)$ is automatic from
\eqref{eq:fL} but the $2\pi$-periodicity is not seen in $f(\th)$ shown
in Fig.~\ref{fig:th-vs-f-24x48-f-df}.

The subvolume method is equivalent to modifying the value of $\th$
inside the subvolume.
If the difference of $\th$ is a multiple of $2 \pi$ and the calculation
respects the $2\pi$-periodicity, the free energy would scale as the
surface area of the subvolume when the subvolume is large enough.
The lack of $2\pi$ periodicity in the free energy density should thus be
interpreted as the presence of a meta-stable vacuum for a fixed value of
$\theta$ (except for $\theta = \pi$ where two vacua interchanged by CP
are degenerate and stable).
Thus, we expect that the meta-stable vacuum should eventually decay into
the stable one by the creation of a dynamical domain wall that attaches
to the interface.

The absence of the decay of the domain into the domain-wall in the
lattice calculation has an analog in the calculation of the static
potential~\footnote{The similar reasoning is found for 2d \cpn-model in
\cite{KeithHynes:2008rw}.}.
The static potential is calculated by inserting a Wilson loop, and
should show the string breaking for configurations with light dynamical
quarks when the two test charges are distant enough.
But it does not occur, at least, within naive methods, and the resulting
potential sticks to the original branch even after passing the
transition point.
The probable reason is that the overlap between the original state with two
static charges and another lower energy state with two mesons are
extremely small.
We infer that the same happens to the calculation of $f(\th)$ for
$\th>\pi$ and the first order phase transition is missing~\footnote{We
expect that the subvolume method can capture second order transitions in
principle because meta-stable states do not exist.}.
It is clearly interesting to directly see the formation of the domain
wall on the lattice though it would not be straightforward.

\begin{figure}[tb]
 \centering
  \includegraphics[width=0.5 \textwidth]
  {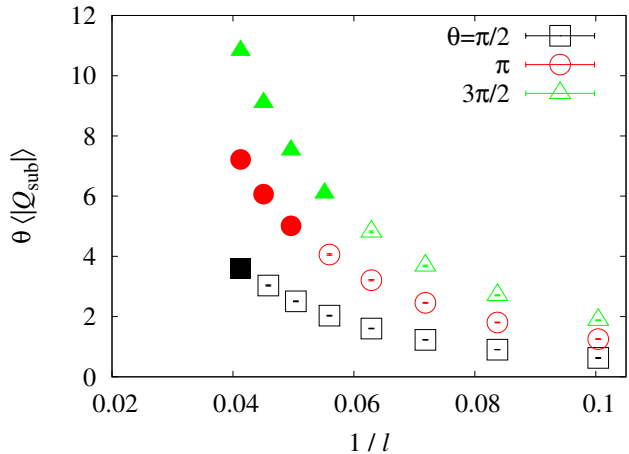}
 \caption{
 Subvolume dependence of $\th\la |\qsub| \ra$.
 The open symbols represent the calculations which succeeded while the
 filled ones represent those which failed.
 }
 \label{fig:allowed-range}
\end{figure}

We have mentioned the $\th$ dependent upper limit on the size of $\vsub$
in sec.~\ref{sec:method}.
We examine the relation between the limit and
$\th\langle|\qsub|\rangle$ at $T=0$.
Figure~\ref{fig:allowed-range} shows $\th\langle|\qsub|\rangle$ as a
function of $1/l$, where we have used the approximate relation
$\langle|\qsub|\rangle=(l^4/V_{\rm full})^{1/2}\langle|Q|\rangle$ and
the measured value of $\langle|Q|\rangle$ at $T=0$ by ignoring the
corrections to the relation of $O(1/l)$ and $O(1/(\chi\vsub))$.
In the figure, the filled symbols represent the points where the
calculations are failed due to the sign problem.
It is seen that the upper limit indeed decreases with $\th$.
Numerical investigation suggests that the upper limit on the subvolume
scales as $1/\th$.
%

\section{Summary}
\label{sec:summary}

We developed the subvolume method, which enables us to extract the $\th$
dependence of the free energy density in 4d Yang-Mills theory not
restricted to $\theta\sim 0$.
At high temperature, the method yields $\th$ dependence consistent with
the instanton prediction, as expected, within the large uncertainty due
to the ambiguity of the scaling region.
To fix this ambiguity, it is necessary to go to larger lattices.
On the other hand, at zero temperature the sign problem arises instead,
but still $f(\th)$ could be calculated to $\th\sim\pi$ with the
systematic uncertainty under control.
Combining the numerical result with the theoretical requirement leads to
the conclusion that the vacuum of 4d \SU(2) Yang-Mills theory undergoes
spontaneous CP violation at $\th=\pi$ as large $N$ theory does.
Although the overlap problem prohibits the domain-wall from being
formed, it is interesting to learn that such a object actually exists in
the Yang-Mills theory~\cite{Luscher:1978rn}.

We have tested the stability of the results by exchanging the order of
extrapolations and obtained the consistent results with enlarged
uncertainties.

In order to promote this study to the quantitative level, it is
necessary to perform lattice simulations with larger volumes and
finer lattice spacings.
Further studies will be presented in the forthcoming
paper~\cite{InProgress}.
It is a fascinating question to ask if our method is applicable for
other questions with sign problems, such as gauge theories with finite
values of the chemical potential.

While numerical results are not accurate past $\th \sim 3 \pi/2$, 
there are indications that the derivative $df(\theta)/d\theta$ decreases
past $\theta=\pi$, and becomes smaller near $\theta \sim 2\pi$.
This is consistent with the expectation \cite{Yamazaki:2017ulc} that
there are two meta-stable branches of the \SU(2) theory, each of which
has $4\pi$ periodicity.

\section*{Acknowledgments}

This work is based in part on the Bridge++ code~\cite{Ueda:2014rya}
and is supported in part by JSPS KAKENHI Grant-in-Aid
for Scientific Research (Nos.~19H00689 [RK, NY, MY], 18K03662 [NY],
19K03820, 20H05850, 20H05860 [MY]) and MEXT KAKENHI Grant-in-Aid for Scientific Research on
Innovative Areas (No.~18H05542 [RK]).
Numerical computation in this work was carried out in part on the
Oakforest-PACS and Cygnus under Multidisciplinary Cooperative Research
Program (No.~17a15) in Center for Computational Sciences, University of
Tsukuba; Fujitsu PRIMERGY CX600M1/CX1640M1 (Oakforest-PACS) in the
Information Technology Center, the University of Tokyo.


\end{document}